\documentclass[aps,prc,groupedaddress,showpacs,twocolumn,floatfix]{revtex4}
\usepackage{graphicx}

\def\deg{^\circ}
\def\gtorder{\mathrel{\raise.3ex\hbox{$>$}\mkern-14mu
 \lower0.6ex\hbox{$\sim$}}}
\def\ltorder{\mathrel{\raise.3ex\hbox{$<$}\mkern-14mu
 \lower0.6ex\hbox{$\sim$}}}
\def\mugegm{\mu_p G_{E_p} / G_{M_p}}
\def\gegm{G_{E_p} / G_{M_p}}
\def\ge{G_{E_p}}
\def\gm{G_{M_p}}

\begin{document}

\title{How well do we know the electromagnetic form factors of the proton?}

\author{J. Arrington}

\affiliation{Argonne National Laboratory, Argonne, IL 60439}

\date{\today}

\begin{abstract}

Several experiments have extracted proton electromagnetic form factors from
elastic cross section measurements using the Rosenbluth technique. Global
analyses of these measurements indicate approximate scaling of the electric
and magnetic form factors ($\mugegm \approx 1$), in contrast to recent
polarization transfer measurements from Jefferson Lab. We present here a
global reanalysis of the cross section data aimed at understanding the
disagreement between the Rosenbluth extraction and the polarization transfer
data. We find that the individual cross section measurements are
self-consistent, and that the new global analysis yields results that are still
inconsistent with polarization measurements. This discrepancy indicates a
fundamental problem in one of the two techniques, or a significant error in
polarization transfer or cross section measurements.  An error in the
polarization data would imply a large error in the extracted electric form
factor, while an error in the cross sections implies an uncertainty in the
extracted form factors, even if the form factor {\it ratio} is measured
exactly.

\end{abstract}

\maketitle


\section{INTRODUCTION}

The electromagnetic structure of the proton is described by the electric and
magnetic form factors. Over the past several decades, a large number of
experiments have measured elastic electron-proton scattering cross sections in
order to extract the electric and magnetic form factors, $\ge(Q^2)$ and
$\gm(Q^2)$ (where $-Q^2$ is the four-momentum transfer squared), using the
Rosenbluth technique~\cite{rosenbluth50}. The electric and magnetic form
factors have been extracted up to $Q^2 \approx 7$~GeV$^2$ by direct Rosenbluth
separations, and these measurements indicate approximate form factor scaling,
{\it i.e.,} $\mugegm \approx 1$ (where $\mu_p$ is the magnetic dipole moment of
the proton), though with large uncertainties in $\ge$ at the highest $Q^2$
values~\cite{walker94, walkerphd}.

\begin{figure}
\includegraphics[height=6.0cm,width=8.0cm]{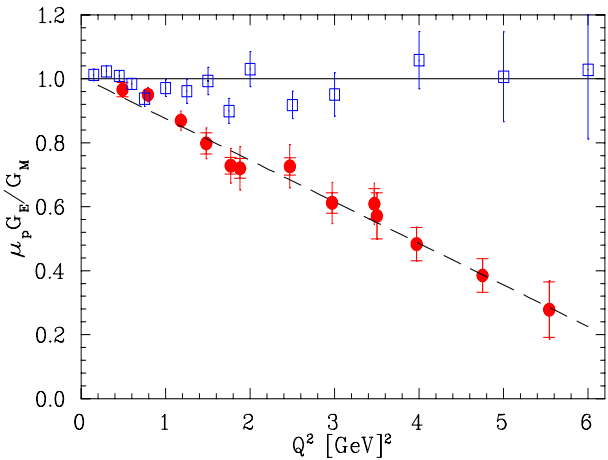}
\caption{Ratio of electric to magnetic form factor as extracted by Rosenbluth
measurements (hollow squares) and from the JLab measurements of recoil
polarization (solid circles).  The dashed line is the fit to the polarization
transfer data.\label{fig:gegm}}
\end{figure}

More recently, elastic electron-proton polarization transfer measurements have
been performed to obtain the ratio $\gegm$. A low $Q^2$ measurement at
MIT-Bates~\cite{milbrath99, milbrath98} obtained values of $\gegm$ consistent
with previous Rosenbluth separations.  Later experiments at Jefferson Lab
(JLab) extended these measurements up to $Q^2 = 5.6$~GeV$^2$~\cite{jones98,
gayou01, gayou02}, and show significant deviations from form factor scaling.
They show a roughly linear decrease of the value of $\mugegm$ from unity at
low $Q^2$ to approximately 0.3 at $Q^2 = 5.6$~GeV$^2$. Figure~\ref{fig:gegm} shows
the JLab polarization transfer measurements from refs.~\cite{jones98,
gayou02}, along with a global Rosenbluth analysis of the cross section
measurements~\cite{walker94}. While the polarization transfer technique allows
much better measurements at high $Q^2$ values, there is a significant
discrepancy even in the region where both techniques have comparable
uncertainties.

When we combine the cross sections with polarization transfer measurements
to extract the form factors (see Sect.~\ref{sec:xsecplushalla}, or
Ref.~\cite{brash02}), we find that the values obtained for $\ge$ are
significantly different, while $\gm$ differs only at the few percent level,
compared to extractions that use only the cross sections. Clearly, it is
necessary to understand the discrepancy in the extracted ratio before we can be
confident in our knowledge of $\ge$ and, to a lesser extent, $\gm$. The
Rosenbluth data is more sensitive to systematic uncertainties, and it has been
suggested that the different Rosenbluth extractions are inconsistent, and
thus unreliable. We will examine the consistency of the Rosenbluth
measurements to test this suggestion. However, even if it is demonstrated that
the cross sections going into the Rosenbluth extractions were incorrect, it
would not completely solve the problem. The polarization measurements
determine only the ratio of $\ge$ to $\gm$, and so reliable cross sections are
still needed to extract the actual values of the form factors. Finally, if the
discrepancy arises from a fundamental problem with either of these techniques,
it may have implications for other measurements.

The goal of this analysis it to better understand the discrepancy between
the Rosenbluth and polarization transfer results.  We begin by demonstrating
that the individual Rosenbluth measurements yield consistent results when
analyzed independently, so that the normalization uncertainties between
different measurements do not impact the result.  We then perform a global
analysis of the cross section measurements, and determine that the results
cannot be made to agree with the polarization results by excluding an small
set of measurements, or by making reasonable modifications to the relative
normalization of the various experiments. The paper is organized as follows:
In section~\ref{sec:review}, we will review the two techniques and summarize
the current measurements of the form factors. In section~\ref{sec:reanalysis},
we present a new Rosenbluth analysis of the cross section measurements, and
compare this to the polarization transfer results and examine possible
scenarios that might explain the discrepancy between the techniques, such as
problems with individual data sets or improper treatment of normalization
uncertainties when combining cross sections from different experiments. In
section~\ref{sec:discussion} we will discuss the results of the analysis and
implications of the discrepancy between the two techniques. Finally, in
section~\ref{sec:summary}, we summarize the results and discuss further tests
that can be performed to help explain the disagreement between the techniques.

\section{OVERVIEW OF FORM FACTOR MEASUREMENTS}\label{sec:review}

We begin with a brief description of the Rosenbluth separation and recoil
polarization techniques, focusing on the existing data and potential problems
with the extraction techniques.

\subsection{Rosenbluth Technique}

The unpolarized differential cross section for elastic scattering can be
written in terms of the cross section for scattering from a point charge and
the electric and magnetic form factors:
\begin{equation}
\frac{d\sigma}{d\Omega} = \sigma_{\rm Mott}
\biggl[ \frac{\ge^2 + \tau \gm^2}{1+\tau} + 2\tau \gm^2 \tan^2{(\theta_e/2)} \biggr],
\end{equation}
where $\theta_e$ is the electron scattering angle, and $\tau=Q^2/4M_p^2$. One
can then define a reduced cross section,
\begin{equation}
\sigma_R \equiv \frac{d\sigma}{d\Omega} \frac{\epsilon(1+\tau)}{\sigma_{\rm Mott}}
= \tau \gm^2(Q^2) + \epsilon \ge^2(Q^2),
\end{equation}
where $\epsilon$ is the longitudinal polarization of the virtual photon
($\epsilon^{-1} = 1+2(1+\tau)\tan^2{(\theta_e/2)}$). At fixed $Q^2$, {\it
i.e.,} fixed $\tau$, the form factors are constant and $\sigma_R$ depends
only on $\epsilon$.  A Rosenbluth, or longitudinal-transverse (L-T), separation
involves measuring cross sections at several different beam energies while 
varying the scattering angle to keep $Q^2$ fixed while varying $\epsilon$.
$\ge^2$ can then be extracted from the slope of the reduced cross section
versus $\epsilon$, and $\tau \gm^2$ from the intercept. Note that because the
$\gm^2$ term has a weighting of $\tau / \epsilon$ with respect to the $\ge^2$
term, the relative contribution of the electric form factor is suppressed at
high $Q^2$, even for $\epsilon=1$.

Because the electric form is extracted from the difference of reduced cross
section measurements at various $\epsilon$ values, the uncertainty in the
extracted value of $\ge^2(Q^2)$ is roughly the uncertainty in that difference,
magnified by factors of $(\Delta\epsilon)^{-1}$ and $(\tau \gm^2 / \ge^2)$.
This enhancement of the experimental uncertainties can become quite large
when the range of $\epsilon$ values covered is small or when $\tau$ ($=
Q^2/4M_p^2$) is large. This is especially important when one combines
high-$\epsilon$ data from one experiment with low-$\epsilon$ data from another
to extract the $\epsilon$-dependence of the cross section. In this case,
an error in the normalization between the data sets will lead to an error in
$\ge^2$ for all $Q^2$ values where the data are combined. If $\mu_p \ge =
\gm$, $\ge$ contributes at most 8.3\% (4.3\%) to the total cross section at
$Q^2=5 (10)$~GeV$^2$, so a normalization difference of 1\% between a
high-$\epsilon$ and low-$\epsilon$ measurement would change the ratio
$\mugegm$ by 12\% at $Q^2=5$~GeV$^2$ and 23\% at $Q^2=10$~GeV$^2$, more if
$\Delta\epsilon < 1$.  Therefore, it is vital that one properly account for
the uncertainty in the relative normalization of the data sets when extracting
the form factor ratios. The decreasing sensitivity to $\ge$ at large $Q^2$
values limits the range of applicability of Rosenbluth extractions; this was
the original motivation for the polarization transfer measurements, whose
sensitivity does not decrease as rapidly with $Q^2$.

\subsection{Recoil Polarization Technique}\label{sec:poltrans}

In polarized elastic electron-proton scattering, $p(\vec{e},e^\prime \vec{p})$,
the longitudinal ($P_l$) and transverse ($P_t$) components of the recoil
polarization are sensitive to different combinations of the electric and
magnetic elastic form factors. The ratio of the form factors, $\gegm$, can be
directly related to the components of the recoil polarization~\cite{dombey69,
akheizer74, arnold81, e93027prop}:
\begin{equation}
\frac{\ge}{\gm} = - \frac{P_t}{P_l}
\frac{(E_e+E_{e^\prime})\tan{(\theta_e/2)}}{2M_p},
\end{equation}
where $E_e$ and $E_e^\prime$ are the incoming and scattered electron energies,
and $P_l$ and $P_t$ are the longitudinal and transverse components of the
final proton polarization.
Because $\gegm$ is proportional to the ratio of polarization components,
the measurement does not require an accurate knowledge of the beam polarization
or analyzing power of the recoil polarimeter. Estimates of radiative
corrections indicate that the effects on the recoil polarizations are
small and at least partially cancel in the ratio of the two polarization
component~\cite{afanasev01}.

\begin{figure}
\includegraphics[height=6.0cm,width=8.0cm]{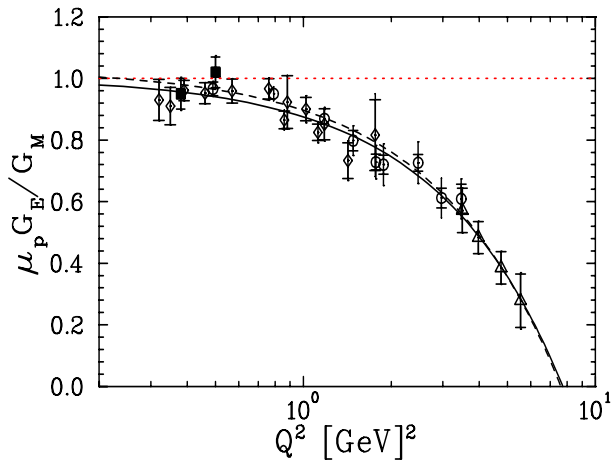}
\protect{\caption{Ratio of electric to magnetic form factor as extracted by recoil
polarization measurements at MIT-Bates (solid squares) and JLab (hollow
symbols). The solid line is the fit to~\cite{jones98, gayou02} (diamonds and
triangles) from Ref.~\cite{gayou02}, while the dashed line is a combined fit,
including the systematic uncertainties (assumed to be fully correlated within
each data set).\label{fig:gegm_poltrans}}}
\end{figure}

Figure~\ref{fig:gegm_poltrans} shows the measured values of $\mugegm$ from the
MIT-Bates~\cite{milbrath99, milbrath98} and JLab~\cite{jones98, gayou01,
gayou02} experiments, both coincidence and single arm measurements, along with
the linear fit of Ref.~\cite{gayou02} to the data from~\cite{jones98, gayou02}:
\begin{equation}
\mugegm = 1 - 0.13(Q^2-0.04),
\label{eqn:ratiofit}
\end{equation}
with $Q^2$ in~GeV$^2$. Comparing the data to the fit, the total $\chi^2$ is is
34.9 for 28 points, including statistical errors only. Assuming that the
systematic uncertainties for each experiment are fully correlated, we can vary
the systematic offset for each data set and the total $\chi^2$ decreases to
33.6. If we allow the systematic offset to vary for each data set {\it and}
refit the $Q^2$-dependence to all four data sets using the same
two-parameter fit as above, {\it i.e.,}
\begin{equation}
\mugegm = 1 - \alpha (Q^2-Q_0^2),
\label{eqn:ratiofitnew}
\end{equation}
for $Q^2 > Q_0^2$, unity for $Q^2 < Q_0^2$.
We obtain $\alpha=0.135\pm0.008$, $Q_0^2=0.24\pm0.08$, and $\chi^2$=28.1 for
26 degrees of freedom (dashed line in Fig.~\ref{fig:gegm_poltrans}). The
systematic offsets are small (consistent with zero) for all of the data sets
except the low $Q^2$ JLab measurement~\cite{jones98}, which is increased by
nearly the full (correlated) systematic uncertainty. This fit not only has a
better $\chi^2$, but also decreases the deviation from unity at very low $Q^2$
values, which improves the agreement with the very precise Rosenbluth results
available below $Q^2 = 1.0$~GeV$^2$.

The main systematic uncertainties come from inelastic background processes and
determination of the spin-precession, both of which have been carefully
studied and accounted for in the JLab measurements.  While this technique
should be less sensitive to systematic uncertainties than the Rosenbluth
extractions, the discrepancy appears at relatively low $Q^2$ values, where
both techniques give equally precise results. Because almost all of the
polarization transfer data comes from the same experimental setup, it is in
principle possible that an unaccounted for systematic error could cause a
false $Q^2$-dependence in the ratio. There are no known problems or
inconsistencies in these measurements and this technique.  At this time, there
is no explanation for the different results obtained by the two techniques. If
we do not understand this discrepancy, then it is difficult to know how to
correctly combine the polarization transfer measurements with the cross section
measurements in order to extract the individual form factors.

\section{Reanalysis of the Rosenbluth Measurements}\label{sec:reanalysis}

The global Rosenbluth analysis shown in Fig.~\ref{fig:gegm} may disagree with
the polarization transfer results for a variety of reasons: inclusion of bad
data points or data sets in the fit, or improper constraints on the relative
normalization of data sets. To better understand the discrepancy, we have
performed a reanalysis of the Rosenbluth measurements.  We will use this to
look for errors or inconsistencies in the data sets and to test
possible explanations for the discrepancy between the two techniques.

An initial analysis reproduced the results of the previous global
fit~\cite{walker94, bosted94}. At this point, several modifications were made
to the data set for subsequent fits: radiative corrections were updated for
some of the older measurements, certain experiments were subdivided into
separate data sets, some normalization uncertainties were updated from those
used by Walker~\cite{walker94}, several cross section measurements were
updated with the final published results, and a set of data points were
excluded.  These modifications are described in detail in the following
section.

\subsection{Data Selection}\label{sec:dataselection}

The global analysis presented here is similar to the one presented in
Refs.~\cite{walker94, walkerphd}. Table~\ref{tab:experiments} shows the data
sets included in the fit, along with a summary of the kinematics for each data
set. The experiments included are, for the most part, the same as in the
previous analysis. Two additional data sets with measurements in the relevant
$Q^2$ region have been included~\cite{stein75, rock92}.  For two of the
experiments included in the previous fit, we use the final published cross
sections~\cite{sill93, andivahis94}, which were not available at the time of
the previous global analysis.

\begin{table}
\caption{Experiments included in global fit, along with the $Q^2$ range, electron scattering
angle range, and normalization uncertainty.\label{tab:experiments}}
\begin{ruledtabular}
\begin{tabular}{lcccc}
Reference			& $Q^2$		& $\theta$ & Norm.	& Laboratory \\
				& [GeV$^2$]	& [degrees]& Uncert.	& 	 \\
\hline
Janssens-1966~\cite{janssens66}	& 0.16-1.17	& 40-145 & 1.6\%	& Mark III	\\
Bartel-1966~\cite{bartel66}	& 0.39-4.09	& 10-25	 & 2.5\%	& DESY		\\
Albrecht-1966~\cite{albrecht66}	& 4.08-7.85	& 47-76	 & 8.0\%	& DESY		\\
Albrecht-1967~\cite{albrecht67}	& 1.95-9.56	& 76	 & 3.0\%	& DESY		\\
Litt-1970~\cite{litt70}		& 1.00-3.75	& 12-41	 & 4.0\%	& SLAC		\\
Goitein-1970$^1$~\cite{goitein70} & 0.27-1.75	& 20	 & 2.0\%	& CEA		\\
Goitein-1970$^1$~\cite{goitein70} & 2.73-5.84	& 19-34	 & 3.8\%	& CEA		\\
Berger-1971~\cite{berger71}	& 0.08-1.95	& 25-111 & 4.0\%	& Bonn		\\
Price-1971~\cite{price71}	& 0.27-1.75	& 60-90	 & 1.9\%	& CEA		\\
Bartel-1973$^2$~\cite{bartel73}	& 0.67-3.00	& 12-18	 & 2.1\%	& DESY		\\
Bartel-1973$^2$~\cite{bartel73}	& 0.67-3.00	& 86	 & 2.1\%	& DESY		\\
Bartel-1973$^2$~\cite{bartel73}	& 1.17-3.00	& 86-90	 & 2.1\%	& DESY		\\
Kirk-1973~\cite{kirk73}		& 1.00-9.98	& 12-18	 & 4.0\%	& SLAC		\\
Stein-1975~\cite{stein75}	& 0.10-1.85	& 4	 & 2.4\%	& SLAC		\\
Bosted-1990~\cite{bosted90}	& 0.49-1.75	& 180	 & 2.3\%	& SLAC		\\
Rock-1992~\cite{rock92}		& 2.50-10.0	& 10	 & 4.1\%	& SLAC		\\
Sill-1993~\cite{sill93}		& 2.88-31.2	& 21-33	 & 3.0\%	& SLAC		\\
Walker-1994$^3$~\cite{walker94}	& 1.00-3.00	& 12-46	 & 1.9\%	& SLAC		\\
Andivahis-1994$^4$~\cite{andivahis94} & 1.75-7.00	& 13-90	 & 1.8\%	& SLAC 		\\
Andivahis-1994$^5$~\cite{andivahis94} & 1.75-8.83	& 90	 & 2.7\%	& SLAC 		\\
\hline
\multicolumn{5}{l}{$^1$ Split into two data sets (see text).} \\
\multicolumn{5}{l}{$^2$ Split into three data sets (see text).} \\
\multicolumn{5}{l}{$^3$ Data below $20\deg$ excluded.} \\
\multicolumn{5}{l}{$^4$ 8 GeV spectrometer.} \\
\multicolumn{5}{l}{$^5$ 1.6 GeV spectrometer.} \\
\end{tabular}
\end{ruledtabular}
\end{table}

For each data set included in the fit, an overall normalization or scale
uncertainty was determined, separate from the point-to-point systematic
uncertainties. This normalization uncertainty is given in, or was estimated
from, the original publication of the data. In most cases, we use the same
scale uncertainty as in the previous global analysis. For six~\cite{bartel66,
albrecht66, albrecht67, goitein70, price71, sill93} of the sixteen
experiments, the published uncertainties included the normalization
uncertainties.  In the previous fit, these uncertainties were double counted
when an {\it additional} normalization uncertainty was added. For these
experiments, we apply the same normalization uncertainty, but remove it from
the published (total) uncertainties to obtain the point-to-point uncertainties.

Two of the experiments~\cite{bartel73, andivahis94} included data taken with
more than one detector.  There will therefore be different normalization
factor for the data taken in the different detectors. In
Ref.~\cite{andivahis94}, these normalization factors were measured by taking
data at identical kinematics for two $Q^2$ points.  We split the experiment
into two data sets, and fit the normalization factor for each one
independently. This will allow the normalization factor to be determined from
both these direct measurements and the comparison to the full data set.
Because we do not apply the normalization factor determined from the original
analysis, we add a 2\% normalization uncertainty (in quadrature) to the 1.77\%
uncertainty quoted in the original analysis. While this may underestimate the
uncertainty in the normalization, the result would tend to be a larger cross
section for this low-$\epsilon$ data set, which would lead to a smaller value
for $\gegm$. As will be shown, even with this possible bias towards lower
values of $\gegm$, the resulting ratio is clearly higher then the polarization
transfer results. Similarly, the elastic cross sections in
Ref.~\cite{bartel73} include three different sets of data:  electrons
detected in a small angle spectrometer, electrons detected in a large angle
spectrometer, and protons detected in the small angle spectrometer
(corresponding to large angle electron scattering). We divide this experiment
into three data sets, each with its own normalization factor.  Finally,
after an initial analysis, it was observed that the data from
Ref.~\cite{goitein70} was taken under very different conditions for
forward and backward angles (see Sect.~\ref{sec:consistency}), and so this
experiment was also subdivided into two data sets. Thus, the 16 experiments
yield a total of 20 independent data sets for this analysis.

The radiative corrections applied to several of the older
experiments~\cite{janssens66, bartel66, albrecht66, albrecht67, litt70,
goitein70, berger71, price71, bartel73, kirk73} neglected higher order terms.
For the combined analysis of old and new experiments, the Schwinger term
and the additional corrections for vacuum polarization contributions from muon
and quark loops have been included, following Eqns.(A5)-(A7) of
Ref.~\cite{walker94}. These terms have very little $\epsilon$-dependence, and
so do not have a significant effect on direct extractions of $\gegm$ from a
single data set. However, they can modify the $Q^2$-dependence at the 1--2\%
level, which has a small effect when determining the relative normalization of
the data sets.

For some of the older experiments, there are further improvements that could
be made to the radiative corrections, but there is not always enough
information provided to recalculate the corrections using more modern
prescriptions. For these experiments we included only the terms mentioned
above, which were not included in the earlier radiative corrections, and
assume that the stated uncertainties for the radiative correction procedures
are adequate to allow for the generally small differences in the older
corrections. For a few of the earliest experiments, the quoted uncertainties
for the radiative corrections were unrealistically small: total uncertainties
of $<$1\% or small normalization uncertainties only. To verify that this
underestimate of the radiative correction uncertainties does not influence the
final results, we included a 1.5\% point-to-point and a 1.5\% normalization
uncertainty for radiative corrections and repeated the global fits presented
in the following sections. In most cases, this error was small or negligible
compared to the other errors quoted, though for three
experiments~\cite{janssens66, litt70, goitein70}, the 1.5\% uncertainty had a
noticeable impact on either the scale or point-to-point uncertainties. The
additional uncertainties did not noticeably change the result of any of the
fits: the extracted value of $\gegm$ changed by less than 1\% for all of the
fits discussed in the later sections. Note that the results presented in this
paper do not apply this additional uncertainty.

Finally, we excluded the small angle data from Ref.~\cite{walker94}. In our
initial analysis, we saw a clear deviation of this small-angle ($< 15
\deg$) data from our global fits, with or without the inclusion of the
polarization transfer data (see Sect.~\ref{sec:consistency} or
Fig.~\ref{fig:walkerangle}).  The deviation is due to a correction determined
to be necessary for small scattering angles in the analysis of
NE11~\cite{andivahis94} that was not applied to the earlier Walker
data~\cite{walker_priv}. Because there is an identified error in this data,
and because it is not straightforward to apply this correction to the published
results, these small angle data ($\theta < 20\deg$) are excluded from the
analysis.

\subsection{Consistency Checks - Single Experiment Extractions}

\begin{figure}
\includegraphics[height=6.0cm,width=8.0cm]{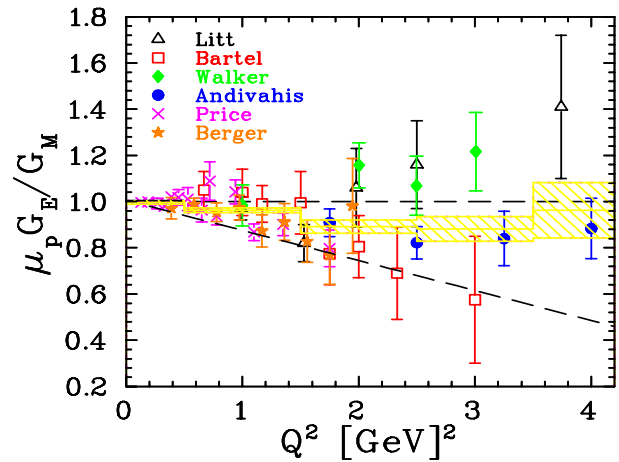}
\caption{Ratio of electric to magnetic form factor from published Rosenbluth
extractions~\cite{litt70,bartel73,walker94,andivahis94,price71,berger71}. The
data are binned into five $Q^2$ bins, and the solid lines (shaded regions)
represent the average (one-sigma range) for the measurement in each bin.  The
dashed lines indicate form factor scaling and the fit to the polarization
transfer data.
\label{fig:gegm_lt_raw}}
\end{figure}

We start by considering the values of $\mugegm$ from published Rosenbluth
extractions of the form factors~\cite{litt70, berger71, price71, bartel73,
walker94, andivahis94} (Fig.~\ref{fig:gegm_lt_raw}).  These are the same
experiments shown in Ref.~\cite{jones98}, where the scatter is used to
illustrate the difficulty of extracting $\ge$ from Rosenbluth separations at
high $Q^2$. The data have been divided into five bins in $Q^2$: 0--0.5 GeV$^2$,
and 1 GeV$^2$ bins above 0.5~GeV$^2$. The solid line shows the weighted
average of all measurements in a given bin while the dotted lines show the one
standard deviation range for each bin. As there is little $Q^2$ variation,
averaging the values in small $Q^2$ bins should give a reasonable measure of
the consistency of the data sets. While the average of these extractions is in
good agreement with the global analysis, there is a large scatter in the
extracted ratios for $Q^2 \gtorder 2$~GeV$^2$. For these measurements, $\chi^2
= 50.6$ for 40 degrees of freedom (45 data points minus 5 fit parameters, the
mean values in each of the 5 $Q^2$ bins), yielding a $\chi^2$ per degree of
freedom, $\chi^2_\nu$, of 1.26 (13\% confidence level (CL)). The agreement is
worse for the higher $Q^2$ data: $\chi^2_\nu = 1.63$ for 17 degrees of freedom
excluding data below $Q^2=1.5$~GeV$^2$ (4.9\% CL). The extent of the scatter
has been used to argue that the Rosenbluth extractions do not give
reproducible results.  The question of the consistency of the data sets must
be addressed before we can draw meaningful conclusions from a global analysis.

While single experiment extractions avoid uncertainties arising from the
relative normalization between different experiments, it should be noted that
most of the form factor ratios shown in Fig.~\ref{fig:gegm_lt_raw} do {\it not}
correspond to single experiment extractions.  Three of these six
extractions~\cite{litt70,berger71,price71} combine new cross section
measurements with cross sections from one or more older experiments.  In the
extraction by Litt~\cite{litt70} the new data are combined with results from
three other experiments. While they give estimates of the effect of a small
change in normalization, the quoted extractions of $\ge$ and $\gm$ ignore the
normalization uncertainties.  The extractions of Berger~\cite{berger71} and
Price~\cite{price71} determine normalization factors between their data and
previous experiments by comparing the cross sections from the different
kinematics to the cross sections calculated assuming the dipole form for both
$\ge$ and $\gm$, in effect assuming that form factor scaling is valid when
determining the relative normalization factors. They do not apply any
uncertainty associated with the determination of these normalization factors.

Two of the six extractions~\cite{bartel73, andivahis94} use data from single
experiments, but use different detectors to measure the large and small angle
scattering. Bartel~\cite{bartel73} does not determine a normalization factor,
but quotes a 1.5\% relative uncertainty between the small angle and large
angle spectrometer data. Andivahis~\cite{andivahis94} determined the relative
normalization factor using data taken at identical kinematics for the SLAC 1.6
GeV and 8 GeV spectrometers.  Unlike the cases where a normalization factor
between two different experiments is determined, no assumption about the
$\epsilon$-dependence goes into the determination.  The uncertainty on the
determination of the normalization factor was applied to the 1.6 GeV
spectrometer which provided a single, low-$\epsilon$ point for each $Q^2$
value measured. However, the uncertainty related to the normalization (0.7\%
for Andivahis, 1.5\% for Bartel) is common to all points, and will have an
effect on $\gegm$ that increases approximately linearly with $Q^2$ (for $Q^2
\gtorder 2$~GeV$^2$). For the Andivahis measurement, this is roughly one-half
the size of the total error, and so the entire data set could move up or down
by roughly half of the total uncertainty shown in the figure.  For the Bartel
data, the high-$Q^2$ points all shift up or down by two-thirds of the total
error due to a one-sigma shift in the normalization factor.

Only one of the experiments~\cite{walker94} used a single detector for both
small and large angle scattering, and is therefore free from normalization
uncertainties. However, this is the experiment for which there was a correction
for the small angle scattering that was not included in the analysis
(Sect.~\ref{sec:dataselection}). 

To study the consistency of the Rosenbluth measurements without the additional
uncertainty caused by combining different experiments, one must examine only
those experiments where an adequate range of $\epsilon$ was covered with a
single detector in a single experiment.  Five of the data sets from
Table~\ref{tab:experiments} cover an adequate range in $\epsilon$ to perform a
single experiment L-T separation~\cite{walker94, litt70, berger71,
andivahis94, janssens66} .  For these experiments, we have used only the cross
section results from the primary measurement, with the updated radiative
corrections as discussed in Sect.~\ref{sec:dataselection}.

\begin{figure}
\includegraphics[height=6.0cm,width=8.0cm]{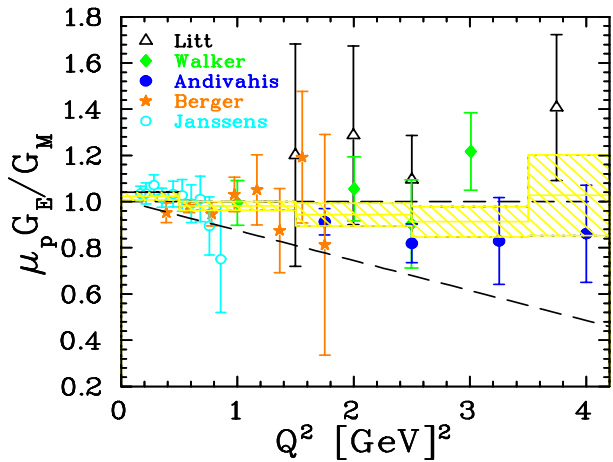}
\caption{Ratio of electric to magnetic form factor as extracted by the
Rosenbluth technique, including only data sets where both forward and
backward angle data were taken in the same apparatus
(Table~\ref{tab:directlt_values}).  The solid lines (shaded regions) indicate
the average (one-sigma range) for the measurements in each bin. The dashed
lines indicate form factor scaling and the fit to the polarization transfer
data.  The Janssens data has 19 $Q^2$ points, which have been rebinned to 10
data points.
\label{fig:gegm_1111111}}
\end{figure}

Table~\ref{tab:directlt_values} shows the values of $\ge / G_{\rm dipole}$, 
$\gm / G_{\rm dipole}$, and $\mugegm$ from the reanalysis of the experiments
that had adequate $\epsilon$ coverage.  Higher order terms in the radiative
corrections have been applied to those experiments which did not include
these terms, as discussed in Section~\ref{sec:dataselection}.

\def\baselinestretch{1.0}
\begin{table}
\caption{$\ge / G_{\rm dipole}$, $\gm / G_{\rm dipole}$, and $\mugegm$ as extracted from the
individual data sets shown in figure~\ref{fig:gegm_1111111}.  A dipole mass of 0.71
GeV$^2$ is assumed for $G_{\rm dipole}$.
\label{tab:directlt_values}}
\begin{ruledtabular}
\begin{tabular}{lcccc}
Data Set		&$Q^2$		& $\ge/G_{\rm dip}$	& $\gm/G_{\rm dip}$	& $\mugegm$	\\
			&(GeV$^2$)	&		&		& 		\\
\hline
Litt
& 1.499	& 1.180$\pm$0.339	& 0.982$\pm$0.111	& 1.201$\pm$0.481 \\
\cite{litt70}
& 1.998	& 1.258$\pm$0.286	& 0.977$\pm$0.072	& 1.287$\pm$0.387 \\
& 2.500	& 1.106$\pm$0.164	& 1.010$\pm$0.028	& 1.095$\pm$0.192 \\
& 3.745	& 1.377$\pm$0.256	& 0.979$\pm$0.038	& 1.407$\pm$0.316 \\

Walker
& 1.000	& 1.006$\pm$0.072	& 1.012$\pm$0.026	& 0.994$\pm$0.097 \\
\cite{walker94}$^1$
& 2.003	& 1.084$\pm$0.120	& 1.027$\pm$0.022	& 1.055$\pm$0.139 \\
& 2.497	& 0.944$\pm$0.180	& 1.045$\pm$0.022	& 0.903$\pm$0.191 \\
& 3.007	& 1.227$\pm$0.145	& 1.008$\pm$0.020	& 1.217$\pm$0.168 \\

Andivahis
& 1.75	& 0.959$\pm$0.053	& 1.049$\pm$0.009	& 0.913$\pm$0.057 \\
\cite{andivahis94}$^2$
& 2.50	& 0.863$\pm$0.082	& 1.054$\pm$0.008	& 0.819$\pm$0.084 \\
& 3.25	& 0.868$\pm$0.185	& 1.047$\pm$0.015	& 0.829$\pm$0.188 \\
& 4.00	& 0.890$\pm$0.205	& 1.033$\pm$0.015	& 0.861$\pm$0.210 \\
& 5.00	& 0.578$\pm$0.453	& 1.030$\pm$0.016	& 0.561$\pm$0.448 \\

Berger
& 0.389	& 0.938$\pm$0.025	& 0.985$\pm$0.019	& 0.952$\pm$0.043 \\
\cite{berger71}
& 0.584	& 0.965$\pm$0.019	& 0.985$\pm$0.009	& 0.980$\pm$0.027 \\
& 0.779	& 0.950$\pm$0.041	& 1.004$\pm$0.012	& 0.946$\pm$0.051 \\
& 0.973	& 1.034$\pm$0.058	& 1.003$\pm$0.016	& 1.031$\pm$0.074 \\
& 1.168	& 1.074$\pm$0.132	& 1.022$\pm$0.023	& 1.051$\pm$0.152 \\
& 1.363	& 0.907$\pm$0.171	& 1.037$\pm$0.022	& 0.875$\pm$0.182 \\
& 1.557	& 1.229$\pm$0.265	& 1.031$\pm$0.027	& 1.192$\pm$0.285 \\
& 1.752	& 0.863$\pm$0.479	& 1.062$\pm$0.036	& 0.813$\pm$0.478 \\

Janssens
& 0.156 & 1.021$\pm$0.028	& 0.926$\pm$0.027	& 1.103$\pm$0.057 \\
\cite{janssens66}
& 0.179	& 0.962$\pm$0.024	& 0.959$\pm$0.016	& 1.003$\pm$0.039 \\
& 0.195	& 0.973$\pm$0.041	& 0.999$\pm$0.032	& 0.974$\pm$0.067 \\
& 0.234	& 1.020$\pm$0.034	& 0.939$\pm$0.025	& 1.087$\pm$0.061 \\
& 0.273	& 1.000$\pm$0.039	& 0.935$\pm$0.019	& 1.070$\pm$0.059 \\
& 0.292	& 1.005$\pm$0.044	& 0.936$\pm$0.022	& 1.074$\pm$0.068 \\
& 0.311	& 0.935$\pm$0.041	& 0.961$\pm$0.018	& 0.974$\pm$0.057 \\
& 0.389	& 1.014$\pm$0.041	& 0.956$\pm$0.016	& 1.061$\pm$0.058 \\
& 0.428	& 1.019$\pm$0.064	& 0.970$\pm$0.024	& 1.051$\pm$0.086 \\
& 0.467	& 0.993$\pm$0.055	& 0.974$\pm$0.020	& 1.020$\pm$0.073 \\
& 0.506	& 1.023$\pm$0.080	& 0.954$\pm$0.029	& 1.073$\pm$0.113 \\
& 0.545	& 0.984$\pm$0.069	& 0.983$\pm$0.020	& 1.000$\pm$0.087 \\
& 0.584	& 1.016$\pm$0.103	& 0.981$\pm$0.030	& 1.036$\pm$0.133 \\
& 0.623	& 0.951$\pm$0.085	& 0.987$\pm$0.020	& 0.964$\pm$0.103 \\
& 0.662	& 0.869$\pm$0.151	& 1.027$\pm$0.031	& 0.846$\pm$0.169 \\
& 0.701	& 1.076$\pm$0.100	& 0.982$\pm$0.021	& 1.096$\pm$0.121 \\
& 0.740	& 1.053$\pm$0.162	& 1.017$\pm$0.032	& 1.036$\pm$0.186 \\
& 0.779	& 0.805$\pm$0.160	& 1.035$\pm$0.022	& 0.778$\pm$0.169 \\
& 0.857	& 0.814$\pm$0.236	& 1.083$\pm$0.023	& 0.751$\pm$0.230 \\

\hline
\multicolumn{5}{l}{$^1$ Data below 15 degrees excluded.} \\
\multicolumn{5}{l}{$^2$ Using data from the 8 GeV spectrometer only.}\\
\end{tabular}
\end{ruledtabular}
\end{table}
\def\baselinestretch{1.5}

The form factors from the single experiment extractions are given in
Table~\ref{tab:directlt_values}, and the form factor ratios are plotted
in Figure~\ref{fig:gegm_1111111}. The total $\chi^2$ is 18.2 for 25 degrees of
freedom (d.o.f.). If the data below $Q^2=1.5$~GeV$^2$ are excluded, $\chi^2 =
10.3$ for 9 d.o.f. (33\% CL). So, while the {\it published} extractions of the
form factors from the different experiments have large scatter and yield
somewhat inconsistent results, it is in part a result of the treatment of
normalization factors in these extractions.  The raw cross sections do not
show this inconsistency, and the true single experiment extractions are
consistent, and agree well with the global Rosenbluth analysis.
Table~\ref{tab:directlts} shows the $\chi^2$ value for each of these data sets
compared to the new global fit to the cross section data
(Fig.~\ref{fig:fits}), and the polarization transfer parameterization of
Ref.~\cite{gayou02}, {\it i.e.,} Eqn.~\ref{eqn:ratiofit}. Every data set
except Janssens, which is limited to $Q^2<1$~GeV$^2$), is in significantly
better agreement with the global Rosenbluth analysis.

\def\baselinestretch{1.0}
\begin{table}
\caption{$\chi^2$ of single experiment extractions compared to $\gegm$ from global
Rosenbluth analysis and polarization transfer ($0.6 < Q^2 < 6.0$).  Bartel is not
included in the sum, as it is not taken from a single data set.
\label{tab:directlts}}
\begin{ruledtabular}
\begin{tabular}{lccc}
Data Set			& data  	& $\chi^2$ (CL) vs. & $\chi^2$ (CL) vs. \\
				& points	& global Rosenbluth & polarization	\\
\hline
Litt~\cite{litt70}		& 4		&	5.39 (24.9\%)	& 15.2 (0.43\%) \\
Walker~\cite{walker94}		& 4		&	5.56 (23.4\%)	& 20.7 (0.04\%) \\
Andivahis~\cite{andivahis94}	& 5		&	1.44 (92\%)	& 13.5 (1.9\%) \\
Berger~\cite{berger71}		& 6		&	2.91 (82\%)	& 8.55 (20\%) \\
Janssens~\cite{janssens66}	& 6		&	3.74 (71\%)	& 4.00 (68\%)  \\
\hline
Sum				& 25		&	19.1 (79\%)	& 62.0 (0.006\%) \\
\hline
Bartel~\cite{bartel73}		& 8		&	6.35 (61\%)	& 10.6 (23\%)  \\
\end{tabular}
\end{ruledtabular}
\end{table}
\def\baselinestretch{1.5}

We could include more data sets if we also used experiments where the
forward and backward angle data are taken with different detectors, but
where the normalization uncertainty is taken into account.
If the Bartel and full Andivahis data sets are included, the results are still
consistent ($\chi^2$=17.4 for 15 d.o.f. for $Q^2 > 1.5$~GeV$^2$),
and the average form factor ratio is again slightly below unity for the
intermediate $Q^2$ values. However, because the uncertainty related to the
normalization is common to all $Q^2$ values, the Bartel data does not strongly
favor either scaling or the polarization transfer result, and the reduction in
the error for Andivahis when both spectrometers are combined is largely offset
by the introduction of the correlated error.

Because of the reduced data set and limited $\epsilon$ range caused by
examining only single experiment extractions, the uncertainties are larger
than for a global analysis. However, this data set should be free from the
additional systematics related to the cross-experiment normalization. The
extracted form factor ratio is slightly below scaling, in good agreement with
the previous global analysis and in significant disagreement with the JLab
polarization transfer results.  In the next stage, we perform a global
analysis of all of the cross section measurements to obtain the most precise
result, and to test possible explanations of the discrepancy between the
Rosenbluth and polarization measurements.

\subsection{Fitting Procedure and Results}\label{sec:fitting}

For the global fits to the cross section data, the form factors are
parameterized using the same form as Refs.~\cite{bosted94,brash02}:
\begin{equation}
G_E(q), G_M(q)/\mu_p = 1~/~[ 1 + p_1 q + p_2 q^2...+p_N q^N],
\label{eqn:fitform}
\end{equation}
where $q = \sqrt{Q^2}$, and $N$ varies between 4 and 8; N=5 is adequate for a
good fit. Because we wish to focus on the discrepancy at intermediate $Q^2$
values, we exclude data below $Q^2=0.6$~GeV$^2$, where the two techniques are
in good agreement, and above $Q^2=8$~GeV$^2$, where the data do not allow for
a Rosenbluth separation. Several parameterizations were tried, and this form
was chosen because it provides enough flexibility to fit the $Q^2$-dependence
of the form factors. This parameterization has odd powers of q, and so will
not have the correct $Q^2 \rightarrow 0$ behavior, but this is not relevant
for this analysis, as we are focusing on the intermediate $Q^2$ range.

\begin{figure}[htb]
\includegraphics[height=6.0cm,width=8.0cm]{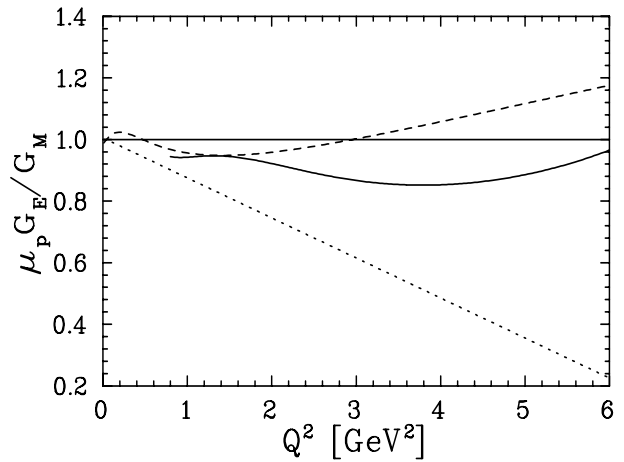}
\protect{\caption{Ratio of electric to magnetic form factor from the
new global fit (solid line), the Bosted fit~\cite{bosted94} (dashed line), and
the fit to the polarization data (dotted line)~\cite{gayou02}.}\label{fig:fits}}
\end{figure}

In addition to the $2N$ parameters for the form factors, we also fit a
normalization factor for each of the data sets.  After partitioning the data
sets for experiments that make multiple independent measurements, there are 20
normalization constants for the 16 experiments, as described in
Sect.~\ref{sec:dataselection}. The fit parameters are allowed to vary, in
order to minimize the $\chi^2_\sigma$, the total $\chi^2$ for the fit to the
cross section data:
\begin{equation}
\chi^2_\sigma = \sum_{i=1}^{N_\sigma} \frac{(\sigma_i - \sigma_{\rm fit})^2}{(d\sigma_i)^2} +
\sum_{j=1}^{N_{\rm exp}} \frac{(\eta_j - 1)^2}{(d\eta_j)^2},
\label{eqn:chisq}
\end{equation}
where $\sigma_i$ and $d\sigma_i$ are the cross section and error for each of
the $N_\sigma$ data points, $\eta_j$ is the fitted normalization factor for
the $j$th data set, and $d\eta_j$ is the normalization uncertainty for that
data set. From the fit, we obtain the values of the fit parameters, $p_i$
(Eqn.~\ref{eqn:fitform}), for both the electric and magnetic form factors, as
well as the normalization factors for each of the 20 data sets.
Figure~\ref{fig:fits} shows the result of our global fit along with the fit of
Bosted~\cite{bosted94} to the previous global analysis and the fit to the
polarization transfer data. While the modifications described in
Sect.~\ref{sec:dataselection} do decrease the $\mugegm$ for $Q^2>2$~GeV$^2$,
the new fit still gives a result that is well above the polarization transfer
result at these momentum transfers. The $\chi^2$ value for this fit , obtained
from Eqn.~\ref{eqn:chisq}, is 162.0 for 198 d.o.f. ({\it i.e.,} $\chi^2_\nu =
0.818$).

\begin{figure}
\includegraphics[height=6.0cm,width=8.0cm]{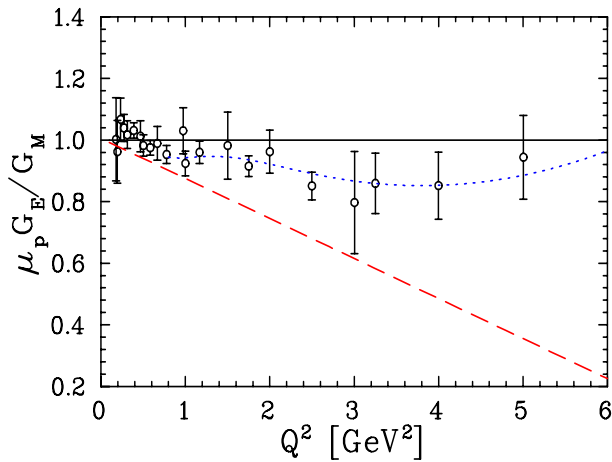}
\protect{\caption{Ratio of electric to magnetic form factor as determined
from direct L-T separations, using the normalizations determined from the
global fit. The dotted line is the ratio from the global fit to $\ge$ and
$\gm$, while the dashed line is the fit to the polarization measurements.}
\label{fig:directlt}}
\end{figure}

In order to estimate the uncertainty of the fit as a function of $Q^2$, we
perform direct L-T separations wherever there are enough data points in a
small range of $Q^2$.  The normalization factors from the global fit are used
to scale each data set, and the $Q^2$-dependence of the fit is used to scale
each point to the central $Q^2$ value. Figure~\ref{fig:directlt} shows the
extracted ratios and uncertainties from these direct L-T separations at 26
$Q^2$ points up to 6 GeV$^2$ (15 above $Q^2$=0.6~GeV$^2$). These $Q^2$ values
are selected by requiring that five or more $\epsilon$ values are in each
$Q^2$ bin, that the $\epsilon$-range of the data is at least 0.6, and that the
correction required to scale the data points to the central $Q^2$ value never
exceeds 2\%. Due to this constraint on the correction, the resulting form
factor ratio is independent of the model used to scale the cross sections for
any reasonable model of the form factors.

To bring the data into agreement with the polarization transfer measurements,
there would have to be a significant $\epsilon$-dependent error in the cross
sections.  Assuming that all of the data sets have such an error, and that it
is linear in $\epsilon$, it would have to introduce an $\epsilon$-dependence
of 5--6\%, nearly independent of $Q^2$, for $Q^2 \gtorder 1.0$~GeV$^2$. This
error would have to be even larger if it affected only some of the data sets,
or if all of the uncertainty came at very large or very small values of
$\epsilon$, {\it e.g.,} if the error only occurred at very small angles or
very low energies.  In addition, a repeat of the above fit, excluding data at
high ($\epsilon > 0.8$) or low ($\epsilon < 0.3$) values of $\epsilon$ did not
have significant impact on the overall fit, changing the extracted values of
$\ge$ by $<$5\% for $Q^2$ values below 4~GeV$^2$.  At higher $Q^2$ values, the
cuts on $\epsilon$ led to larger changes in $\ge$, but the reduction in the
$\epsilon$ range did not allow for a precise extraction of $\ge$, and the
changes were small compared to the final uncertainty.

\subsection{Consistency Checks}\label{sec:consistency}

As a first check of the consistency of the data sets, we examine the
contribution to the total $\chi^2$ coming from each data set. No data set had
an excessively large $\chi^2$ value, only five of the data sets have
$\chi^2_\nu > 1$ and all five of these have reasonable confidence levels
(none below 10\%). In addition, no individual cross section value had
excessively large ($>3\sigma$) deviations from the fit. However, the fact that
the fit gives such a low $\chi^2_\nu$ value, combined with the fact that most
of the individual data sets have $\chi^2_\nu < 1$, indicates that several data
sets may have overly conservative estimates for the uncertainties. Therefore,
the total $\chi^2$ value of the global fits cannot be viewed as an absolute
measure of goodness of fit, and we will focus on the {\it change} in $\chi^2$
between different fits when using the same data sets as a measure of the ${\it
relative}$ goodness of fit.

While these statistical measures help us locate individual data sets that are
inconsistent with the bulk of the data, they are not always enough to detect
systematic errors in the data, especially when the error estimates are
somewhat conservative. It is quite possible that these statistical measures
will overlook systematic errors which are small compared to the individual
uncertainties, but still large enough to modify the small
$\epsilon$-dependence extracted. Thus, we would also like to look for any
systematic trends in the data sets which might modify the result. 

\begin{figure}
\includegraphics[height=5.0cm,width=8.0cm]{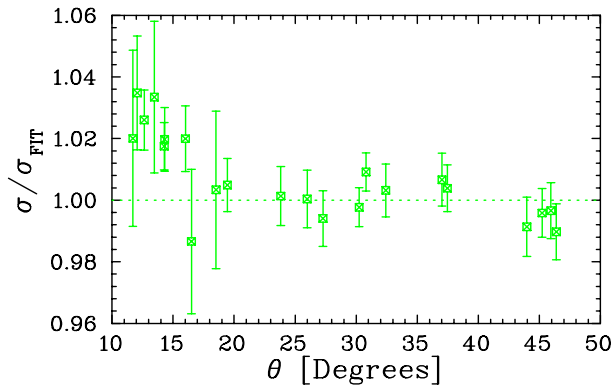}
\protect{\caption{Ratio of cross section to global fit for Walker
data~\cite{walker94}, showing a systematic deviation at low scattering angles.}
\label{fig:walkerangle}}
\end{figure}

For each data set we compare the measured cross sections to the fit and look
for systematic deviations from the global fit as a function of $Q^2$,
$\theta$, $E_e^\prime$ or $\epsilon$. A systematic $\epsilon$-dependence of
just 1--2\% may not show up in the total $\chi^2$ of the data set (especially
if the errors are overly conservative or are noticeably larger than 1--2\%),
but could lead to a noticeable change in the electric form factor at high
$Q^2$ values.  In the preliminary version of the global fit, it was observed
that the Walker data~\cite{walker94} showed a clear deviation from the global
fit at small angles (Fig.~\ref{fig:walkerangle}), even though the $\chi^2$
value for this data set was quite reasonable. The data below 20$\deg$ was then
removed from later fits.

Similarly, the Goitein data~\cite{goitein70} showed a systematic deviation from
the initial global fit.  The ratio of cross section from this data set to that
obtained from the global fit was systematically higher by $\gtorder 5$\% for
the higher $Q^2$ values.  The original publication listed several differences
in the larger $Q^2$ data: different collimation, additional cuts, and several
corrections which were negligible for the low $Q^2$ data were quite large for
these points. Because of the difference in running conditions between low and
high $Q^2$, it was decided to break up this data set into two subsets, each
with its own normalization factor, and to increase the normalization
uncertainty from 2.8\% to 3.8\% for the large $Q^2$ data due to the larger and
less well understood corrections for this data set. The high $Q^2$ data could
also have been excluded altogether, but while the conditions for this subset
of the data were different, there was no clear evidence of any specific
problem.

As another check for systematic trends in the data, we examine the direct L-T
separations performed at several values of $Q^2$. For the separations 
shown in Fig.~\ref{fig:directlt}, we plot the cross sections versus $\epsilon$
to look for systematic deviations from the expected linearity and to look for
data sets which have systematic differences in their $\epsilon$-dependence. 
After separating the Goitein data into a two subsets the only data set which
stands out is the data from Janssens~\cite{janssens66}, which has some
deviations from a linear $\epsilon$-dependence.  Sometimes the higher
$\epsilon$ points are above the extracted (linear) $\epsilon$-dependence,
sometimes they are below, and sometimes they just show non-statistical
scatter.  Because there are no systematic trends to these deviations, and
because there was no indication of problems in the original publication, we do
not exclude this data set from the analysis. The effects of excluding specific
data sets was tested separately, and the results are presented in the
following section.

\subsection{Bad Data Sets?}\label{sec:exclusion}

\begin{figure}
\includegraphics[height=6.0cm,width=8.0cm]{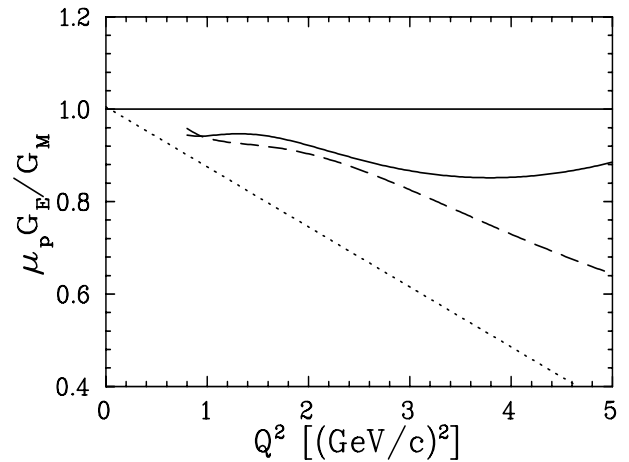}
\protect{\caption{Ratio of electric to magnetic form factor from the
global fit of all of the data sets (solid curve), and the lowest ratio
found by excluding up to three of the data sets (dashed curve).
Also shown is the fit to the polarization data (dotted line).}\label{fig:fits_kill3}}
\end{figure}

Because most of the high $Q^2$ Rosenbluth data comes from a limited set of
experiments, it is possible that a single data set may have an error that
strongly biases the global analysis. The global fit described in
Sect.~\ref{sec:fitting} was repeated 20 times, with a different data set
excluded each time. The exclusion of a single data set generally had little
effect on the global fit, although there were a few data sets whose exclusion
lead to a noticeable increase (up to $\sim 10$\% for $Q^2 > 2$~GeV$^2$) or a
noticeable decrease in $\mugegm$ (from 5\% to 15\% for $3 < Q^2 < 5$~GeV$^2$).
Even excluding two or three data sets together generally had little effect on
the result. The three data sets whose exclusion lead to the greatest decrease
in the ratio were identified, and the fit was repeated with all three
excluded. Figure~\ref{fig:fits_kill3} shows the result of the full global fit,
and the global fit with all three of these data sets removed. There is little
change for $Q^2 < 3$~GeV$^2$, where both techniques have high precision, while
there is a significant decrease for $Q^2 > 3$~GeV$^2$, where the ratio is
constrained by a very limited set of experiments and is very poorly
constrained when the three data sets are removed.  Also, while the ratio at
large $Q^2$ is significantly decreased, it is still well above the polarization
transfer result.

\subsection{Consistency of Global Fit and Polarization Transfer Results}
\label{sec:xsecplushalla}

The new global fit is clearly in disagreement with the polarization transfer
results, and shows no indications of inconsistency between the data sets or
bias due to inclusion of erroneous results. One remaining possibility is that
it is the fitting procedure itself, rather than the data, that leads to this
discrepancy. While we include the relative normalization factors for the
experiments in the analysis and obtain best-fit values for the normalizations,
it is possible that a small change in the normalization factors for some or
all of the experiments would significantly improve the agreement with the
polarization transfer data without dramatically increasing the overall
$\chi^2$ of the fit.

To test this possibility, we performed a constrained fit, using the same data
as in the previous fit, but fitting only $\gm$ and the normalization factors,
while fixing the ratio $\mugegm$ to the fit of the polarization transfer data
(Eqn.~\ref{eqn:ratiofit}). The extracted magnetic form factor is roughly 2\%
higher over the entire $Q^2$ range ($Q^2 \gtorder 1$~GeV$^2$), due to the
reduction in strength from the electric contribution. The constrained fit has
5 more degrees of freedom (the parameters that were used to fit $\ge$) while
the total $\chi^2$ for the fit increases by 40.2 (from $\chi^2$=162.0 for 198
d.o.f. to $\chi^2$=202.2 for 203 d.o.f.).

\begin{figure}
\includegraphics[height=6.0cm,width=8.0cm]{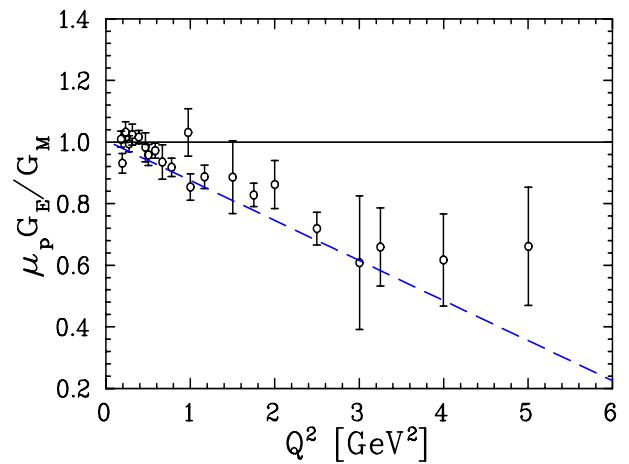}
\protect{\caption{Ratio of electric to magnetic form factor as determined
from direct L-T separations, using the normalization factors determined
by forcing the ratio to reproduce the polarization transfer fit (dashed line).
Although the normalization factors were fitted to reproduce the
polarization transfer results, the direct L-T separation is systematically
higher than the fit at large $Q^2$ values.}\label{fig:directlt_force}}
\end{figure}

Even though the normalization factors are adjusted in order to best reproduce
the polarization transfer results, the direct L-T separation using these
normalization factors is systematically high for all $Q^2$ values, as seen in
figure~\ref{fig:directlt_force}. So not only is this fit significantly
worse, the normalization factors derived from constraining the $\gegm$ ratio
to match polarization transfer data still do not fully explain the magnitude
of the falloff of $\gegm$. The effect is only slightly less using the combined
fit to all four recoil polarization data sets. With the ratio forced to
$\mugegm = 1 - 0.135(Q^2-0.24)$ (Eqn.~\ref{eqn:ratiofitnew}), the total
$\chi^2$ is 197.3, an increase of 35.3 in the total $\chi^2$, with 5 extra
degrees of freedom.

Because of the conservative error estimates on some of the data sets, the
reduced $\chi^2$ is still less than one, even after this noticeable increase.
Therefore, the absolute value of $\chi^2_\nu$ cannot be directly used to
measure goodness of fit.  To get an idea of the relative goodness
of fit, one can scale the overall uncertainties on the cross sections such
that $\chi^2_\nu$ for the unconstrained fit is approximately one. 
This means reducing the cross section uncertainties by a factor of 0.905,
leading to $\chi^2_\nu$=1 for the unconstrained fit to the cross section
data, $\chi^2_\nu$=1.22 (1.9\% CL) for the fit constrained to
Eqn.~\ref{eqn:ratiofit}, and $\chi^2_\nu$=1.19 (3.5\% CL) for the fit
constrained to Eqn.~\ref{eqn:ratiofitnew}.  Assuming that the unconstrained fit
should yield $\chi^2_\nu=1$ is arbitrary, but it is a reasonable staring point
since we expect $\chi^2_\nu \approx 1$ if the estimated uncertainties are
correct and if our fitting function can accurately reproduce the data. One
would expect $\chi^2$ to be even higher, and the confidence levels for
the constrained fits to be even lower, if the fitting function does not
adequately reproduce the data or if there are any inconsistencies in the cross
sections. So it is likely that these confidence levels are upper limits of the
consistency between the cross section data and the parameterizations of the
polarization transfer data.

Forcing the fit to match the parameterization of the polarization transfer
data gives too much weight to polarization data, as it neglects the
uncertainties in the polarization measurements.  To avoid this, we also
performed a combined fit, treating the cross section and polarization transfer
measurements on an equal footing. We repeated the procedure described in
Section~\ref{sec:fitting}, but with the inclusion of the polarization transfer
ratios as additional data points, and with a systematic offset included for
each data set~\cite{jones98, gayou01, gayou02}, as in the fit from
section~\ref{sec:poltrans} (Eqn.~\ref{eqn:ratiofitnew}). The $\chi^2$ for
this combined fit is the contribution from the cross section measurements
(Eqn.~\ref{eqn:chisq}) plus the additional contribution from the polarization
transfer ratio measurements:
\begin{equation}
\chi^2 = \chi^2_\sigma +
\sum_{k=1}^{N_R} \frac{(R_k - R_{\rm fit})^2}{(dR_{\rm stat})^2} +
\sum_{l=1}^{N_{\rm exp}} \frac{(\Delta_l)^2}{(dR_{\rm sys})^2}
\label{eqn:chisq2}
\end{equation}
where $R = \mugegm$, $dR_{\rm stat}$ and $dR_{\rm sys}$ are the statistical and 
systematic uncertainties in $R$, $N_R$ is the total number of polarization
transfer measurements of $R$, $\Delta_l$ is the offset for each data set, and
$N_{\rm exp}$ is the number of polarization transfer data sets. Because the
polarization transfer data are included in the fit, the normalization factors
for the cross section measurements are adjusted to give consistency between
different cross section data sets, as well as consistency with the
polarization measurements of $\mugegm$, much as they are in the constrained
fit.

\begin{figure}
\includegraphics[height=6.0cm,width=8.0cm]{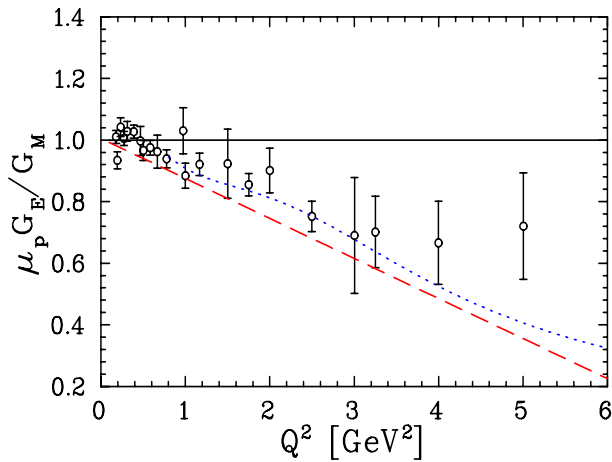}
\protect{\caption{Ratio of electric to magnetic form factor as determined
from combined fit of cross section and polarization transfer data.
The dotted line is the result of the fit, and the circles are the results
from the direct L-T separations using the normalization factors determined
from the global fit.}\label{fig:directlt_551}}
\end{figure}

The ratio of $\gegm$ for this combined fit is systematically higher then
the polarization transfer results (Fig.~\ref{fig:directlt_551}). As with the
constrained fit, the direct L-T separations using the normalization factors
from this global fit are systematically higher than the actual ratio obtained
in the fit, and the quality of fit is significantly reduced when the
polarization transfer data is included: $\chi^2 = 215.8$ for 218 d.o.f.,
an increase in $\chi^2$ of 53.8 for the additional 20 data points.

While the combined fit to cross section and polarization transfer data yields a
ratio that is close to the polarization measurements, this does not imply that
the data sets are yielding consistent results, just that the polarization
transfer results dominate the fit.  The goodness of fit is significantly worse
when the polarization transfer data is added to the cross section
measurements. Because of the conservative errors, it is difficult to estimate
the relative goodness-of-fit for the combined fit.  We can scale the
uncertainties by a factor of 0.905 to force $\chi^2_\nu = 1$ for the fit to
cross sections, yielding $\chi^2_\nu = 1.21$ (1.9\% CL) for the combined fit
to the cross sections and polarization transfer.  However, as discussed for
the constrained fit, this is likely to be an upper limit on the consistency of
the two fits.  The comparison of single experiments L-T separations to the
polarization transfer fit, $\mugegm = 1 - 0.13(Q^2-0.04)$, yields a confidence
level of 0.006\% (Table~\ref{tab:directlts}).  To take into account the
uncertainties of the polarization transfer measurements, we compare the single
experiment L-T separations to the fit with the slope parameter decreased by one
sigma, {\it i.e.,} $\mugegm = 1 - 0.125(Q^2-0.04)$~\cite{brash02}.  This gives
$\chi^2=57.8$ for 25 degrees of freedom, or a 0.02\% confidence level.

We can also use a recently-proposed test to determine the consistency of the
data from the two techniques, following the prescription of
Refs.~\cite{Maltoni:2003cu,Maltoni:2002xd}. They define a test called the
``parameter goodness-of-fit'' (PG) measure, which is designed test the
consistency of independent data sets sharing a common set of parameters. This
approach is significantly less sensitive to over- or underestimates of the
uncertainties of the individual data sets. The corresponding $\chi^2$ for this
test, $\overline{\chi^2}_{\rm min}$, is the increase in $\chi^2$ for a
combined fit, relative to fits of the individual data sets.  The effective
number of degrees of freedom is the number of parameters in common between the
two data sets, which in this case are the parameters that go into the ratio.
So, for comparison of the cross section and polarization transfer results,
$\overline{\chi^2}_{\rm min}$ is the difference between $\chi^2_{\rm min}$ for
the combined fit and the values of $\chi^2_{\rm min}$ for the independent fits
of the cross section data and of the polarization transfer data, {\it i.e.,}
$\overline{\chi^2}_{\rm min} = 215.8 - 162.0 - 26.5 = 27.3$.  Note that the
$\chi^2$ for the fit to polarization transfer data alone is not the result
presented in section~\ref{sec:poltrans}, because the combined fit includes
only the polarization transfer data from JLab~\cite{jones98,gayou01,gayou02}.
The two types of data both constrain the ratio, effectively 5 parameters, and
so the PG goodness-of-fit measure corresponds to $\overline{\chi^2}_{\rm min}
= 27.3$ for 5 degrees of freedom, or a 0.005\% CL. This measure has two
advantages over the confidence levels calculated based on the direct
comparison of the single experiment extractions to the polarization transfer
fits. It includes the entire cross section data set, rather than just the five
single experiments extractions in Table~\ref{tab:directlts}, and it accounts
for the uncertainties of the both the cross section and polarization transfer
measurements.

In the end, both approaches yield a confidence level of $\ll 1\%$, showing that
the data sets are clearly inconsistent. If it is an error in the cross section
data, it would take an $\epsilon$-dependent correction of at least 5--6\% to
make the results from the two techniques consistent.  Until the source of the
inconsistency between the two data sets is determined, we do not know how to
combine the measurements in order to obtain a reliable extraction of the form
factors.

\section{Discussion}\label{sec:discussion}

We have presented a new extraction of the proton electromagnetic form factors
based on a global analysis of elastic cross section data at moderate to high
$Q^2$. The extracted value for $\gegm$ is slightly lower than in the previous
global analysis~\cite{walker94,bosted94}, but is still well above the values
determined by polarization transfer measurements. We have demonstrated that
the discrepancy between the global Rosenbluth analysis and the polarization
transfer results is not merely the result of the inclusion of one or two bad
data sets. While modifying the normalization factors of the different
experiments can lead to a significant reduction in the extracted $\gegm$,
they yield significantly lower quality
fits, and still do not fully resolve the discrepancy between the two
techniques.  Finally, we have demonstrated that the apparent inconsistency
between single experiment extractions is not due to any inconsistency in the
data themselves, but due to improper treatment of normalization uncertainties
when combining data from multiple measurements. The values of $\gegm$ from
single experiment extractions, which avoid the normalization uncertainties
involved in a combined analysis, are self-consistent and in good agreement
with the result of the global analysis.

This indicates that there is a more fundamental reason for the discrepancy,
such as an intrinsic problem with either the Rosenbluth or polarization
transfer techniques, or an error in the cross section or polarization transfer
measurements.  If the error is in the cross section measurements, it must be
a systematic problem that yields a similar $\epsilon$-dependence in a
large set of these measurements: a 5--6\% linear $\epsilon$-dependence in
all cross section measurements above $Q^2 \approx 1$~GeV$^2$, larger if
only some of the data are affected, or if the error has significant deviations
from a linear $\epsilon$-dependence. Since the the cross sections are
necessary to extract the absolute value of $\ge$ and $\gm$ even with very
precise measurements of $\gegm$, an unknown systematic error in the cross
section measurements implies unknown systematic errors in the values of the
form factors.

Until this discrepancy is understood, it is premature to dismiss the
Rosenbluth extractions of the form factors.  There is a significant difference
in the values of $\ge$ extracted by these two techniques, and smaller
($\sim$3\%) differences in the extracted values of $\gm$. In addition to the
impact this uncertainty has on the state of our knowledge of the proton
structure, it also affects other measurements which rely on the proton
electromagnetic form factors as input, or measurements where there are
significant corrections due to the radiative tail of the elastic peak.

While the cross sections extracted using the new polarization transfer ratio
measurements are typically within a few percent of previous parameterizations,
the large change in the {\it ratio} of the form factors means that the changes
in extracted cross sections are strongly $\epsilon$-dependent. Thus, it will
have a significant impact on experiments that try to examine a small
$\epsilon$-dependence.  For example, there will be a large effect on
Rosenbluth separation measurements for $A(e,e')$ ({\it e.g.,} Coulomb sum rule
measurements at large $q$) or $A(e,e'p)$ ({\it e.g.,} $>50$\% difference for
recent results for separated structure functions in
nuclei~\cite{dutta00,dutta_unpub}).

Experiments which need only the {\it unseparated} elastic cross sections
are less sensitive to these uncertainties.  Two such examples are the
extraction of the axial form factors from neutrino scattering measurements,
and extractions of nucleon spectral functions in nuclei, which use the proton
electromagnetic form factors (or elastic cross sections) as input. An analysis
of the neutrino measurements~\cite{budd_priv} shows that the extracted axial
form factor is essentially identical using the Rosenbluth or polarization
transfer parameterizations of the form factors.  The difference also has
relatively little effect on the extraction of the spectral function in
unseparated $A(e,e'p)$ measurements. This is not surprising, as the
polarization measurements only determine the relative contributions from $\ge$
and $\gm$, and the overall size of the form factors is still determined by
fitting the cross section measurements.  With the constraint from the
polarization measurements included, the combined fit can reproduce the
measured cross sections at the few percent level. However, while fits with and
without polarization transfer data can reproduce the measured cross sections
at the few percent level, the discrepancy in the ratio of form factors implies
that some large set of these cross sections may be wrong by 5\% or more. Until
the source of the discrepancy is identified, there is no way to be sure which
of these cross sections are incorrect, or how well these cross sections are
measured.

Finally, it is important to note that the while a {\it consistent} extraction
of $\ge$ and $\gm$ yields only small differences in the elastic cross sections,
this is not always the case when combining extractions of $\ge$ and $\gm$ from
different analyses. Using the polarization transfer data to constrain
$\gegm$ and cross section measurements to determine the size of $\ge$ and
$\gm$, as presented here or in another recent analysis~\cite{brash02}, yields
cross sections that are nearly identical to those from the cross section
analysis only, except for a change of approximately 5\% in their
$\epsilon$-dependence. However, using $\gm$ as extracted from Rosenbluth
extractions, combined with $\gegm$ from polarization transfer measurements can
yield cross sections which are significantly lower than the measured cross
sections.  For example, if one takes the magnetic form factor parameterization
from Bosted~\cite{bosted94}, but calculates the electric form factor using the
polarization transfer ratios, the resulting cross sections are lower by 4--10\%
over a very large $Q^2$ range (0.1--15~GeV$^2$), compared to using both form
factors from the Bosted parameterization. Thus, it is extremely important to
use a consistently determined set of form factors when examining the
difference between the Rosenbluth and polarization transfer data, or when
parameterizing the elastic cross section for comparison to other data or
calculations.

There have been two more recent L-T separations from Hall C at JLab,
which were not included in this analysis.  One took points at $Q^2$=0.64
and 1.81~GeV$^2$~\cite{dutta_unpub}, both of which were in agreement with
the new global Rosenbluth analysis presented here (Fig.~\ref{fig:fits}).  A
more extensive set of Rosenbluth measurements was taken as part of JLab
experiment E94-110~\cite{e94110prop}.  The experiment measured $R =
\sigma_L/\sigma_T$ in the resonance region, but also took elastic data.  This
allowed them to perform several Rosenbluth separations for $0.5 < Q^2 <
5.5$~GeV$^2$~\cite{christy_unpub}.  These results are also in excellent
agreement with the new global fit presented here.  While these newer JLab
measurements have not been included in the new fit, the fact the single
experiment extractions from these measurements are in good agreement with the
global fit indicates that they would not significantly change the final
fit, although their inclusion would decrease the uncertainties in the global
analysis, and thus increase the significance of the discrepancy with the
polarization transfer results.

\section{Summary and Future Outlook}\label{sec:summary}

A careful analysis of Rosenbluth extractions has been performed to test
the consistency of the worlds body of elastic cross section measurements, and
to test explanations for the discrepancy between polarization transfer and
Rosenbluth extractions of the proton form factors.  We find no inconsistency
in the cross section data sets, and cannot remove the discrepancy via
modifications of the relative normalization of different data sets or the
exclusion of individual measurements.

This discrepancy indicates a fundamental problem in one of the two techniques,
or a significant error in either the polarization transfer or cross section
measurements.  An error in the polarization data would imply a large error in
the electric form factor extracted from a combined analysis, and may have
consequences for other recoil polarization measurements. An error in the cross
section data would have to introduce an $\epsilon$-dependence of $>5$\% for
$Q^2 > 1$~GeV$^2$, implying an error in both the electric and magnetic form
factors However, while the uncertainty in the form factors is smaller in this
case, the error in the cross section measurements will also lead to
uncertainties in other measurements which require the elastic cross section as
input.  Thus, even if it is demonstrated that the polarization transfer
measurements are correct, it is necessary to determine the source of the
discrepancy in order to have confidence in our knowledge of the elastic cross
sections, and any other measurement which relies on this knowledge.

Future results from JLab should significantly improve the situation. JLab
experiment E01-001 ran in 2002, and performed a high precision Rosenbluth
separation using a modified experimental technique~\cite{e01001prop,
Arrington:2002cr}.  This experiment should be able to clearly differentiate
between the ratio of $\gegm$ as seen in previous Rosenbluth measurements and
in the polarization transfer results, while being significantly less sensitive
to the types of systematics that are the dominant sources of uncertainties in
previous results.  In addition, there is an approved experiment to extend the
polarization transfer measurements to higher $Q^2$~\cite{e01109prop} using a
different spectrometer from the previous measurements ($Q^2>1$~GeV$^2$), which
all used the High Resolution Spectrometer in Hall A at JLab.  This will
provide the first independent check of the Hall A large $Q^2$ polarization
transfer experiments, whose systematics are dominated by the spin-precession
in the spectrometer. These tests will provide crucial information to help
explain the discrepancy in the measurements of the proton form factors. The
resolution of the discrepancy will significantly improve the state of
knowledge of the proton form factors, as well as determining if other other
measurements utilizing the Rosenbluth or polarization transfer techniques are
affected.

\begin{acknowledgments}

This work is supported by the U. S. Department of Energy, Nuclear Physics
Division, under contract W-31-109-ENG-38.

\end{acknowledgments}

\bibliography{globfit}

\end{document}